# Characteristics of EAS neutron component obtained with PRISMA-32 array

D. M. Gromushkin, F. A. Bogdanov, A. A. Petrukhin, I. I. Yashin, K. O. Yurin
*National Research Nuclear University MEPhI (Moscow Engineering Physics Institute), Kashirskoe shosse 31, 115409 Moscow, Russia*

O. B. Shchegolev, Yu. V. Stenkin, V. I. Stepanov
*Institute for Nuclear Research of RAS, 117312 Moscow, Russia*

The paper is devoted to the results of the EAS neutron component investigations by means of the PRISMA-32 array. The array consists of 32 en-detectors and enables to record delayed thermal neutrons accompanying showers. For registration of thermal neutrons, the scintillator based on $^6$Li isotope as a target is used in the detectors. Some results of the processing of data accumulated over a long period of time are presented: the lateral distribution function of neutrons in EAS and preliminary results on EAS neutron multiplicity spectrum and distribution of showers in e/n ratio.

## 1. INTRODUCTION

PRISMA-32, a novel type of Extensive Air Shower (EAS) array, is operated in the Scientific and Educational Center NEVOD (MEPhI) since 2012 [1]. It was created in cooperation between MEPhI and INR to study neutron component of EAS [2], which is produced in interactions of high-energy hadrons of the shower with nuclei of atoms in the atmosphere and ground surface and provides important information about EAS development. The array allows us to study the thermal and epithermal neutron component using *en-detectors* over a whole area of the array. In addition, electromagnetic component is measured (only for passage of multiple charged particles through the scintillator [3]) by the same detectors. The array layout is shown in figure 1.

An advantage of the thermal neutron component study is that the time profile of EAS in thermal neutrons is of the order of 10 ms, that is about $10^6$ times longer than the time profile of charged particles close to the shower core. This allows us to record the number of neutrons in a wide dynamic range up to 1000 neutrons per a detector per event.

## 2. PRISMA-32 ARRAY

The array consists of two independently operating clusters of 16 en-detectors each and covers an area of about 500 m$^2$. En-detectors are installed inside the experimental building at the 4th floor with a distance of 2.5 or 5 m.

The non-uniform location of the detectors is explained by the existence of a free space in the experimental complex, which contains some other arrays, including Cherenkov water detector (in the center) [4]. Design photo of the en-detector is shown in figure 2. The effective area of each detector is 0.36 m$^2$. Thin layer of inorganic scintillator ZnS (Ag) and LiF (enriched up to 90% of $^6$Li) is used for detecting of EAS neutrons. Two outputs from the 12th and 7th dynodes of FEU-200 photomultiplier are used to expand the dynamic range of electromagnetic component measurements.

A multichannel FADCs (PCI slots) with a frequency of 1 MHz are used to digitize signals from the detectors. Time gate for neutron counting is equal to 20 ms. The pulses are integrated with the time constant of 1 µs using a special discriminator-integrator-unit (DIU) of front-end electronics.

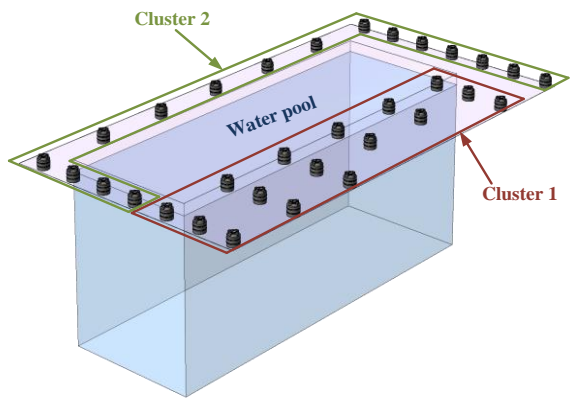

Figure 1: The layout of the PRISMA-32 array.

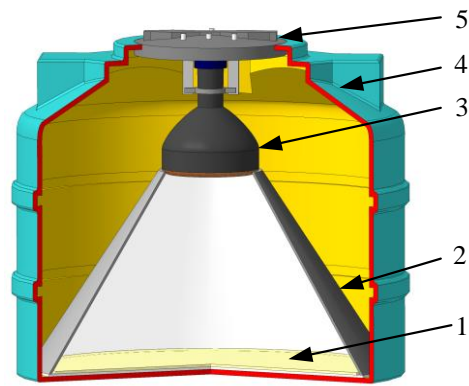

Figure 2: The en-detector design: 1 is ZnS(Ag)+$^6$LiF scintillator; 2 is light reflecting cone; 3 is FEU-200 PMT; 4 is PE water tank; 5 is PE lid.

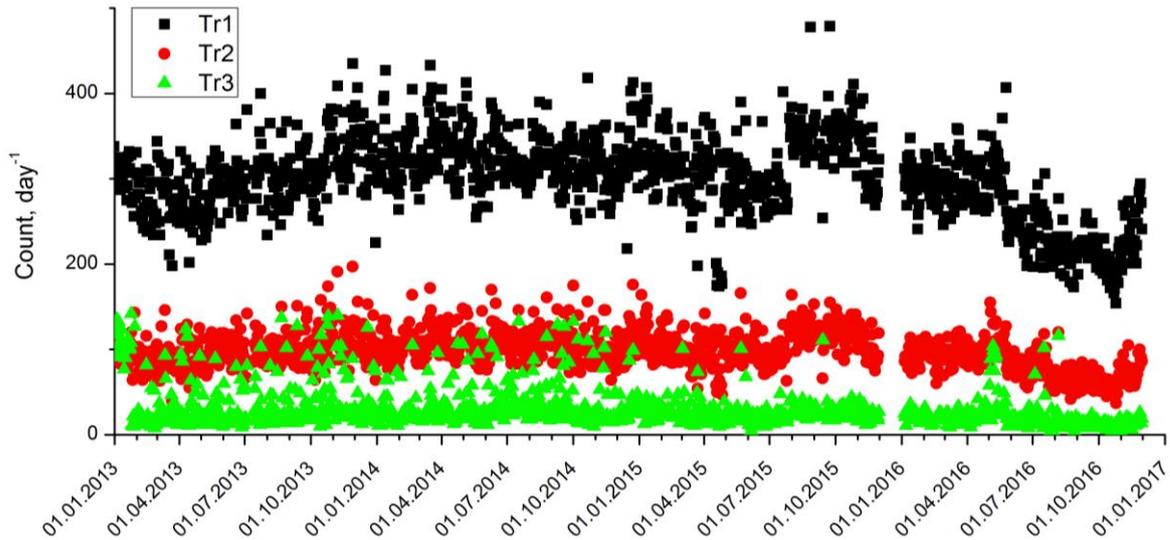

Figure 3: The counting rate of various triggers for the two clusters of PRISMA-32.

The first-level trigger is produced by a coincidence of any two of the 16 detectors of the cluster with a threshold of five particles in a time gate of 1 μs [1]. Cluster structure of the PRISMA-32 detector is presented in figure 4. On-line program analyzes the data and generates a second-level trigger depending on the event type (physical trigger, or marker):

T1, in a case of coincidence of at least two cluster detectors with threshold of about 5 particles (mip) in the first time bin;

T2, in a case when total energy deposit is more than 50 mip in one cluster; and

T3, if the number of neutrons detected in a cluster exceeds 4.

The program also records the energy deposit in each detector, the number of detected neutrons and temporal distribution of these neutrons with a step of 100 μs, and saves this information on a hard disk if there is at least one second-level trigger (counting rate is about 1100 events per day, or 0.013 s$^{-1}$).

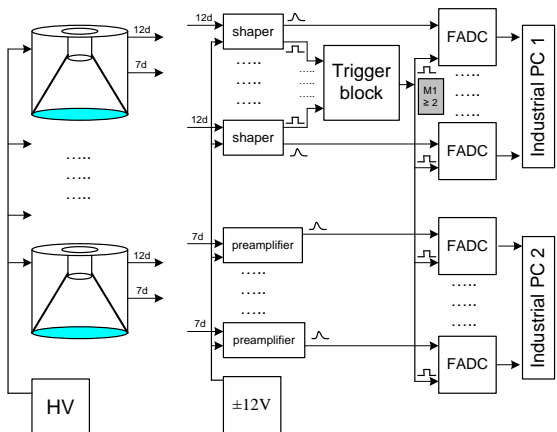

Figure 4: Cluster structure of the PRISMA-32 detector.

## 3. RESULTS

PRISMA-32 array is operated in the continuous mode of data taking. In the present paper, the data on integral distribution in the number of thermal neutrons, lateral distributions of thermal neutrons in EAS, differential and integral distribution of extensive air showers in the ratio of the number of recorded electrons to the number of recorded thermal neutrons obtained for 4.5 years are presented.

Duty time of the array running during this period was about 95%. There are more than 1.5 million event statistics, 10% of which can be attributed to the events of the EAS accompanied by neutrons (operation of more than two detectors, more than 50 charged particles, and the number of neutrons more than 4). Figure 3 shows counting rate of various triggers for the two clusters of PRISMA-32. Throughout the data taking period the array shows a stable performance.

The distribution in the number of recorded neutrons was accumulated with an additional condition: EAS axis should be inside the array area. During 4.5 years there were more than 90 000 events satisfying selection conditions. A standard procedure dealing with the maximum likelihood method to locate EAS axis and to estimate other parameters was used [5]. For these events an integral spectrum in EAS secondary thermal neutrons number was obtained (see figure 5).

As one can see, the integral distribution follows a power law with index β=1.95±0.05. This result is consistent with the spectral slope obtained for the number of hadrons distribution measured by KASCADE [6].

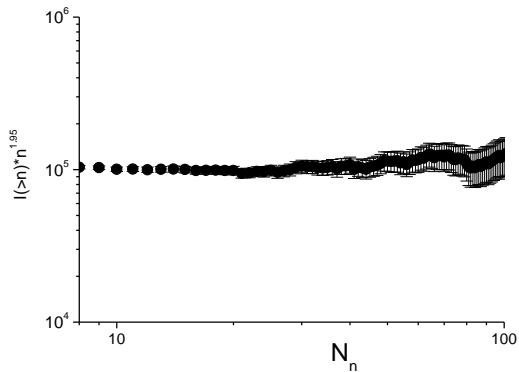

Figure 5: Experimental integral distribution in the number of thermal neutrons.

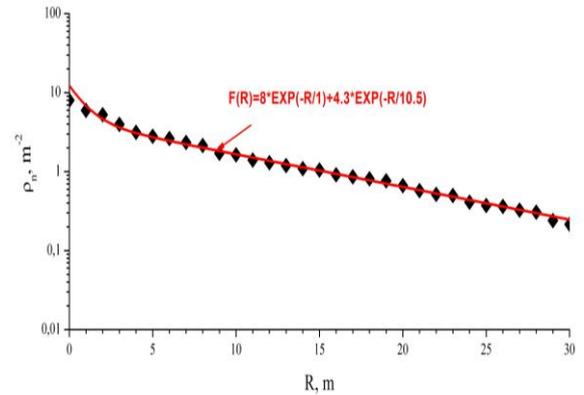

Figure 6: Lateral distribution function of EAS thermal neutrons.

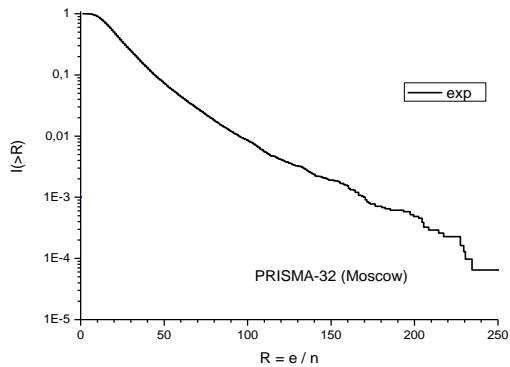

Figure 7: The integral distribution of extensive air showers in the e/n ratio.

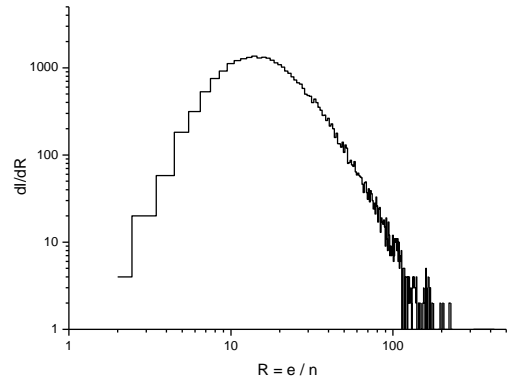

Figure 8: The differential distribution of extensive air showers in the e/n ratio.

The distance from the located core to each detector was found, and lateral distribution of thermal neutrons was obtained (figure 6). From the figure one can see that data can be fitted with double exponential function $y(R) = A_1\exp(-R/r_1) + A_2\exp(-R/r_2) + y_0$ with parameters $r_1=1$ m and $r_2=10.5$ m. The first parameter can be connected with the mean distance of secondary neutrons from parent hadron [7]. The second one ($r_2$) is a typical distance of parent hadrons from the shower core.

Figure 7 and Figure 8 show the distribution in the ratio of the recorded number of shower electrons to the number of recorded thermal neutrons at observation level (e/n) which depends on the primary particles mass [8].

## 4. CONCLUSION

Results of the PRISMA-32 array data analysis for 4.5 years of data taking are presented. Differential and integral distributions of extensive air showers in the ratio of the number of electrons to the number of secondary thermal neutrons are obtained. EAS thermal neutron lateral distribution function as well as EAS size in thermal neutrons in the experimental hall was measured. Obtained results are consistent with our previous data.

## Acknowledgments

The work was performed as a part of the grant of the President of the Russian Federation MK-7597.2016.2. It was also supported by the RFBR grants 16-32-00054 and 16-29-13067.

## References

[1] D.M. Gromushkin et al., Journal of Instrumentation, **9**, C08028, 2014.
[2] D.M. Gromushkin et al., Physics of Atomic Nuclei, **78**, 349, 2015.
[3] D.M. Gromushkin et al. Bull. RAS. Physics, **79**, 380, 2015.
[4] O. Saavedra et al., J. Phys.: Conf. Ser., **409**, 012009, 2013.
[5] A.A. Petrukhin et al., PoS (ICRC2015) 427, 2015; http:pos.sissa.it.
[6] J.R. Hörandel et al., Proc. 27th ICRC, vol. 1, 137, 2001.
[7] O.B. Shchegolev et al., Bull. Lebedev Phys. Inst., **43**, 223, 2016.
[8] Yu.V. Stenkin et al., Bull. RAS. Physics, in press, 2017.